\documentclass[pre,amsmath,amssymb]{revtex4}
\usepackage{graphicx}
\usepackage{dcolumn}
\usepackage{bm}
\usepackage{subfigure}
\usepackage{epstopdf}
\begin{document}

\title{Shape regulation generates elastic interaction between living cells}
\author{Roman Golkov}
\author{Yair Shokef\footnote{\texttt{shokef@tau.ac.il}}}
\affiliation{School of Mechanical Engineering and The Sackler Center for Computational Molecular and Materials Science, Tel Aviv University, Tel Aviv 69978, Israel.}

\begin{abstract}
The organization of live cells to tissues is associated with the mechanical interaction between cells, which is mediated through their elastic environment. We model cells as spherical active force dipoles surrounded by an infinite elastic matrix, and analytically evaluate the interaction energy for different scenarios of their regulatory behavior. We obtain attraction for homeostatic (set point) forces and repulsion for homeostatic displacements. When the translational motion of the cells is regulated, the interaction energy decays with distance as $1/d^4$, while when it is not regulated the energy decays as $1/d^6$. This arises from the same reasons as the van der Waals interaction between induced electric dipoles.
\end{abstract}

\maketitle

\section{Introduction}

\subsection{Mechanobiological background}

Live cells exert contractile forces on their environment \cite{Salbreux2012536}. The elastic behavior of the extracellular matrix (ECM) and its rigidity affect the forces that cells apply and consequently the resultant displacements \cite{Winer2009, Ghibaudo2008, Nisenholz2014, Shenoy2016}, the cell migratory behavior \cite{Reinhart-King2008, Korrf, Shi14012014, Notbohm20150320, Guo10042012}, and cell division \cite{ayelet}. Alterations of the shape and size \cite{Solon2007} and also of the rigidity~\cite{Tee2011} of cells in response to changes of the ECM rigidity were also observed experimentally. For stem cells the rigidity of the ECM may even affect their biological phenotype, from neuronal on soft substrates, muscle on substrates with intermediate rigidity to bone on rigid substrates \cite{Engler2006}. It was also shown experimentally that the interaction behavior between cells is related to the elastic moduli and the non-linear elastic behavior of the substrate \cite{Reinhart-King2008}, and that presence of rigid boundaries affects the behavior of cells \cite{Mohammadi}.

One of the immediate conclusions from these experimental results is that cells sense changes in their mechanical environment and respond to those changes in a variety of ways: by modifying the applied forces and displacements, their size and shape and even their rigidity. This, in turn suggests that there are mechanical interactions between cells through the environment~\cite{He2014116}. Indeed, the dependence of the interaction distance between live cells on the rigidity of the substrate was observed~\cite{Reinhart-King2008}. It was also shown that on nonlinear substrates, cells respond to the presence of each other over relatively long distances (up to 10 cell diameters)~\cite{Winer2009} in contrast to linear substrates, on which the response distance is limited to 1-2 cell diameters. This fact is attributed to the strain-stiffening behavior of the ECM~\cite{Winer2009, Leong2010287} or to fiber buckling \cite{Notbohm20150320}. Improved models of biopolymer gels predict an increase in the range of transmission of internal cellular forces \cite{PhysRevE.92.032728,Rosakis}. An alternative hypothesis is that the fibrous nature of the ECM makes the transmission of mechanical signals to such long distances possible \cite{Rudnicki2013, Ma2013, Abhilash, Ronceray, Wang}.

In this paper we model cells as spherical force dipoles, which create contractile forces on their surface. We seek the interaction energy between two such cells surrounded by an infinite ECM. Realistically, this ECM has nonlinear material properties, thus theoretically analyzing such interactions is a very ambitious goal~\cite{Shokef2012, shokeferr}. Here we lay the conceptual foundations for reaching this goal by focusing on active force dipoles surrounded by a linearly-elastic, or Hookean material. We suggest that the concepts we introduce and the physical mechanisms that we identify would be relevant also to nonlinear media. We define the interaction energy as the additional amount of work performed by the force dipoles as a result of the presence of neighboring dipoles. From studies of the micromechanics of elastic inclusions we know that in the case of two bodies of any shape that apply a hydrostatic pressure on a surrounding linearly elastic material, their interaction energy vanishes~\cite{02, 03}. We define this type of force dipoles as ``dead''. As an example of such behavior one may think of heating a system with inclusions in a medium with a different thermal expansion coefficient. The self-displacements applied by each such inclusion do not depend on the distance to neighboring inclusions. In contrast to this, in this paper we introduce ``live'' behavior as self-regulation of the applied forces or displacements in order to preserve their shape in the presence of the interaction with other cells, or force dipoles. We calculate the interaction energy between two ``live'' force dipoles as a function of the separation distance between them. 

The mechanical environment of the cell may change due to the presence of other cells, due to external forcing, or due to changes in the medium's rigidity, and it is not clear how cells respond to such changes. The working hypothesis of many studies, and which we will adopt here, is that there is some sort of \textit{mechanical homeostasis} in the cell. Namely, the cell tends to maintain certain quantities~\cite{Saez, whycellscare}. For instance, cells may regulate the forces that they apply, and then the displacement that they generate will vary with environmental changes, or alternatively, cells may regulate their deformation, and then the forces required to generate those displacements will vary. \emph{We show that shape regulation of the force dipoles leads to attractive interaction  for force homeostasis and to repulsive interaction for displacement homeostasis.} The interactions we find are analogous to the van der Waals interaction between two induced electric dipoles, thus corroborating the mechanobiological elasticity-electrostatics analogy~\cite{schwarz, PhysRevE.69.021911, RevModPhys.85.1327}.

It is important to emphasize that real cells probably do not keep their exact shape, however we suggest that generically self-regulation related to the interplay between active forces and the cell's shape could generate an interaction which is qualitatively similar to what we find. Our work deals with an abstract model of the cell and its mechanical behavior, which in real life are clearly more complicated. Yet we hope that insights obtained from our analytical solution of this ideal picture will shed light on the understanding of interactions in more realistic scenarios. In particular it would be interesting to relate our work to recent work on the relation between cell morphology and the polarization of the active forces that the cell applies in response to its mechanical environment~\cite{Hakkinen2011, Mousavi2014, Shenoy2016}.

\subsection{Cells as spherical active force dipoles}

\begin{figure}[h]
\centering
\hspace*{\fill}
\subfigure[]{\includegraphics[trim= -0mm -45mm 0mm 0mm,width=0.12\textwidth]{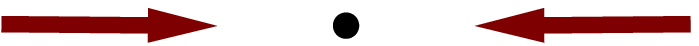}\label{dipole1}} \hfill
\subfigure[]{\includegraphics[trim= 0mm 0mm 0mm 0mm,width=0.18\textwidth]{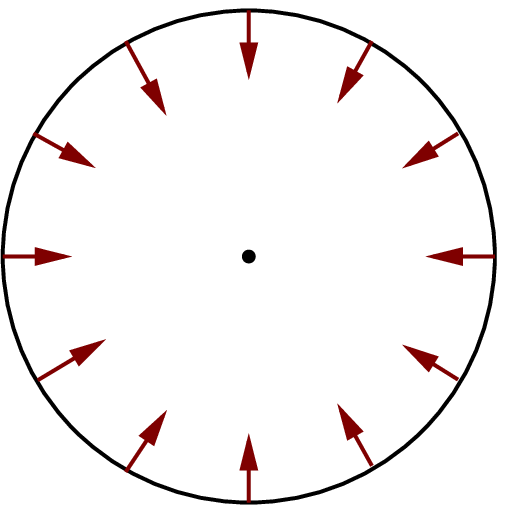}\label{dipole2}}
\hspace*{\fill}
\caption{(a) Linear and (b) spherical force dipoles.} \label{fig_dipoles}
\end{figure}

In analogy to Eshelby's elastic inclusions as force-generating centers~\cite{eshelby1,eshelby2}, the contractile activity of cells can be modeled as contractile force dipoles~\cite{schwarz}, see Fig.~\ref{fig_dipoles}. We adopt this approach, and for simplicity introduce spherical force dipoles as spherical bodies, which apply isotropic contractile forces on their environment \cite{Shokef2012}. The contractile forces at the surface of each spherical force dipole represent the cellular forces that are generated by the contraction of the actin network by myosin motors and are transferred to the ECM trough focal adhesions. By analogy with electric dipoles in electrostatics, a force dipole is defined in mechanics as two equal and opposite point forces applied at some distance from each other, see Fig.~\ref{dipole1}. Our spherical force dipole consists of an infinite number of such linear force dipoles with the same center point, distributed isotropically on the surface of the sphere, see Fig.~\ref{dipole2}.

\subsection{No interaction between ``dead'' spherical force dipoles}

\begin{figure}[h]
\centering
\includegraphics[width=0.6\textwidth]{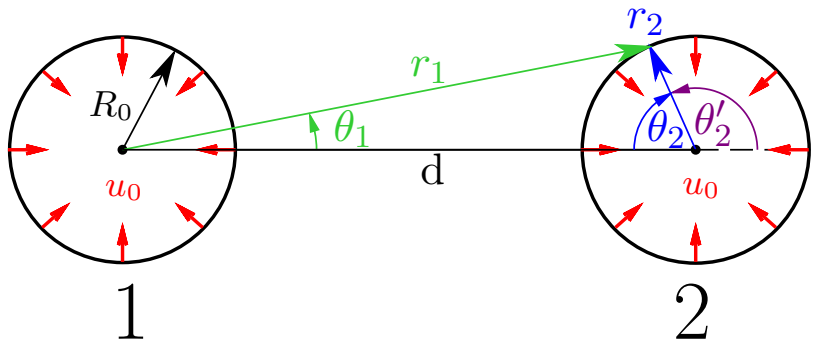}
\caption{Two spherical force dipoles each with radius $R_0$, both applying a radial isotropic displacement $u_0$ on their surfaces. The coordinate system of sphere $1$ is right handed (green) and the coordinate system of sphere $2$ is left handed (blue) and may be written as $\theta_2=\pi-\theta'_2$ where $\theta'_2$ is the commonly-used right-handed azimuthal coordinate for sphere 2.}\label{circles3}
\end{figure}

The interaction energy vanishes for two spherical inclusions in an infinite, linearly-elastic medium, each of which induces symmetric displacements or stresses in the principal directions; namely an initial volume change only with no distortion of their shapes \cite{02,03}, see Fig. \ref{circles3}. This result may be explained in the following way: the dilation of each sphere creates a pure shear field around it and a solely compressive field inside. The energy of interaction between the two spheres may be expressed in terms of the stress from one sphere times the strain from the other sphere, integrated over the interior of the spheres~\cite{03,Simes}. Since one field is a pure shear and the other is pure compression, their coupling does not generate an additional energy. Explicit evaluation of the vanishing interaction energy in this case~\cite{PhysRev.37.1527} is given in Appendix~\ref{appa}. We note that this result holds if the elastic moduli inside the inclusions are identical to those of their environment, while if the elastic moduli differ, the interaction energy will not vanish~\cite{Eshelby1955487}.

\begin{figure}[t]
\centering
\includegraphics[trim=0mm 3mm 0mm 0mm,width=0.45\textwidth]{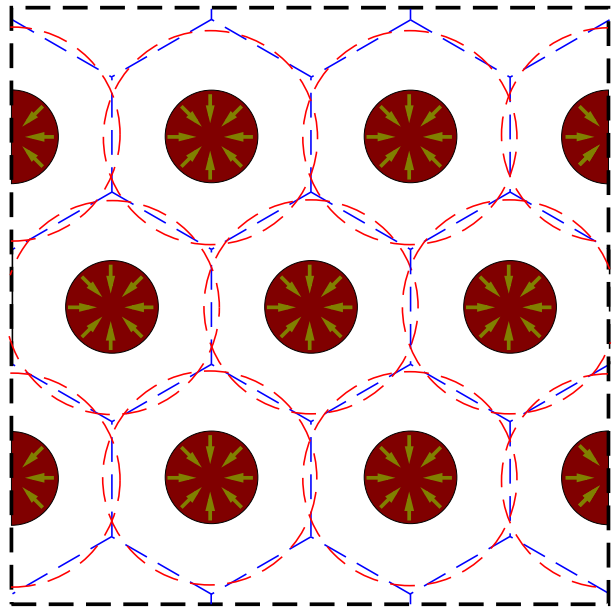}\\
\caption{Triangular lattice of spherical force dipoles, its Wigner-Seitz unit cell (blue dashed line), and approximated spherical unit cell (red dashed line).} \label{WZ}
\end{figure}

For calculating the interaction energy in an infinite periodic array of spherical active force dipoles, we first note that this situation is equivalent to a single force dipole in its Wigner-Seitz unit cell, see Fig. \ref{WZ}. The symmetry of this periodic array dictates that the normal displacement on the midplane between each two neighboring force dipoles must vanish, and thus we may equivalently analyze a spherical force dipole surrounded by elastic ECM limited by a rigid polyhedral Wigner-Seitz unit cell. Since we describe the material as linearly elastic, if we do not take into account regulation in the activity of the force dipoles due to the deformations generated by their neighbors, we can employ the aforementioned result regarding two ``dead'' force dipoles and conclude that also for a system of many such force dipoles the interaction energy vanishes.

\subsection{Spherical Wigner-Seitz unit cell approximation}

Instead of solving the displacement field with the boundary condition of vanishing displacement on the actual boundaries of the polyhedral Wigner-Seitz unit cell, we previously approximated the unit cell as having a spherical shape~\cite{whycellscare}. Within this approximation we could exactly calculate the interaction energy by solving the spherically-symmetric equations of mechanical equilibrium. We define the interaction energy as the difference between the work performed by the spherical force dipole to displace the medium by $u_0$ when it is  surrounded by a rigid spherical Wigner-Seitz unit cell and the work when it is surrounded by an infinite ECM. For a force dipole of radius $R_0$ surrounded by ECM with bulk modulus $K$ and shear modulus $G$. The self-energy of  a single force dipole in an infinite medium is given by (see Section \ref{Eoie} below): $E_0=8 \pi G R_0 u_0^2$.
This is the basic energy scale in our problem, and all interaction energies we obtain scale with $E_0$. For a single force dipole enclosed by a spherical Wigner-Seitz unit cell of radius $R_c$, we relate $R_c$ to the distance between neighboring force dipoles, and define the dimensionless distance as $\tilde{R}_c=R_c/R_0$. The interaction energy in the displacement-homeostasis scenario is given by~\cite{whycellscare}: $E_{int}=  \frac{3-\nu}{4(1-\nu)}  \frac{E_0}{\widetilde{R}_c^3}$\label{Eintinfnd}, where $\nu = \frac{3K-2G}{2(3K+G)} $ is the Poisson ratio of the ECM. This interaction energy represents a repulsive force which decays algebraically with dipole-dipole separation. If instead we consider stress homeostasis and fix the stress $\sigma_0$ that the force dipole applies on the medium, we may write $E_0=\frac{\pi \sigma^2_0 R_0^3}{2 G}$ and now $E_{int}=-\frac{3-\nu}{4(1-\nu)}  \frac{E_0}{\widetilde{R}_c^3}$, representing an attractive interaction.

\subsection{Motivation}

\begin{figure}[t]\label{two}
\centering
\subfigure[``Dead'']{\fbox{\includegraphics[width=0.465\textwidth]{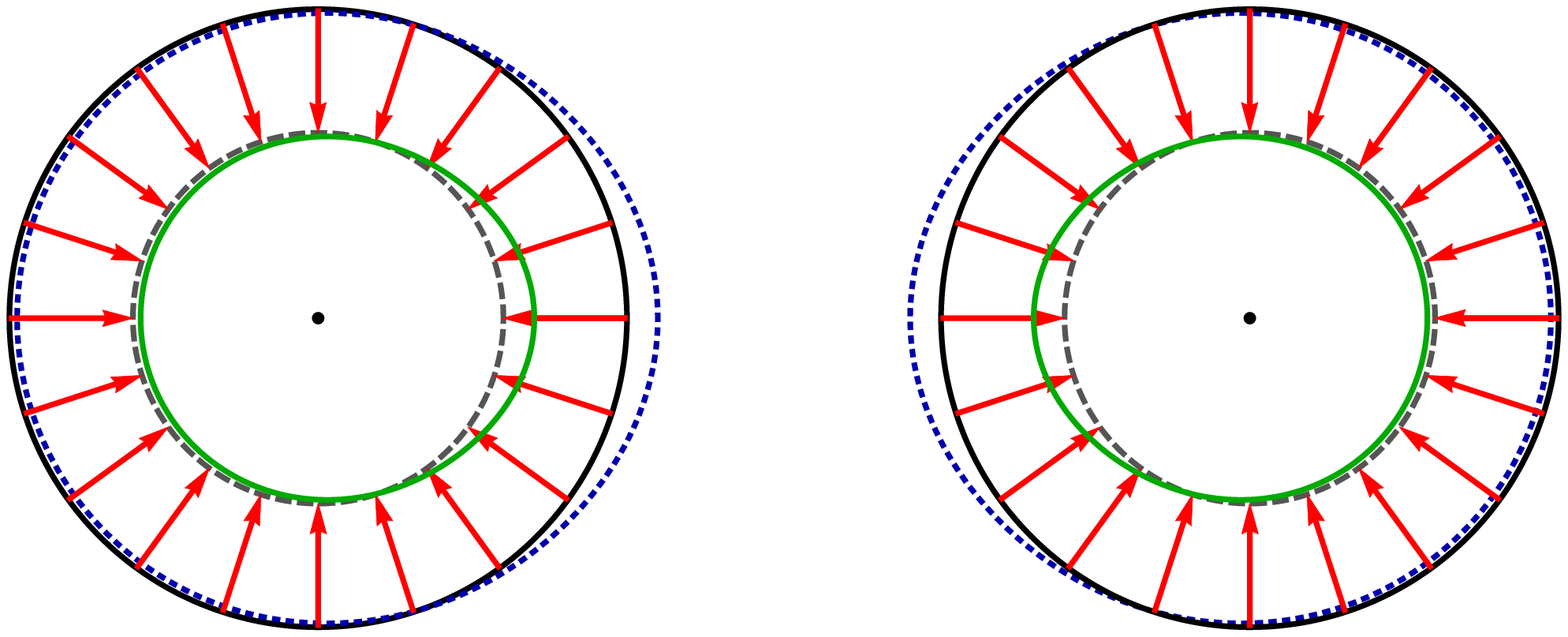}\label{twoa}}} \\
\subfigure[``Live'', fixed-size, fixed-position]{\fbox{\includegraphics[width=0.465\textwidth]{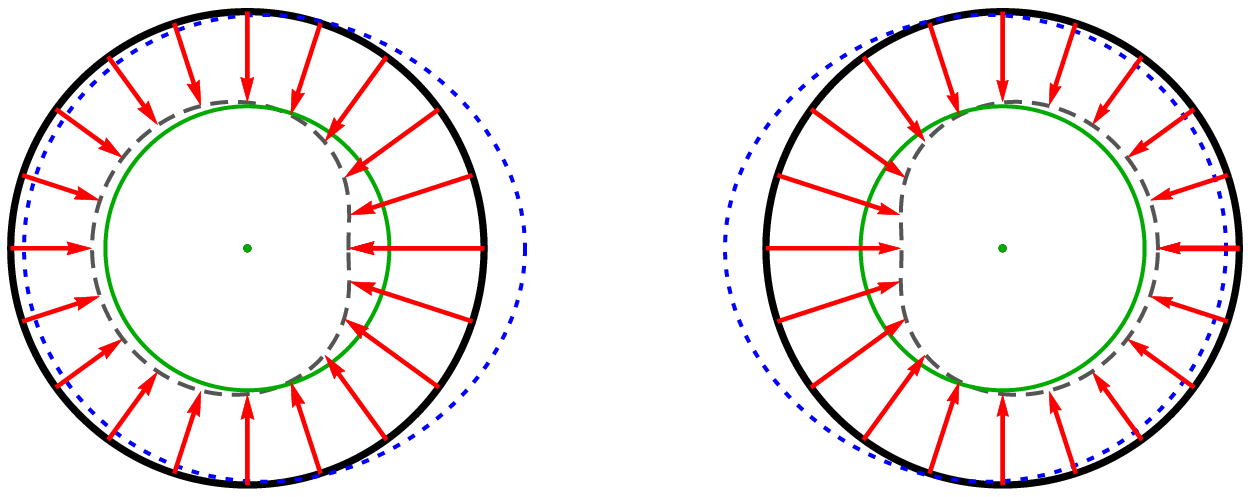}\label{twob}}} \\
\subfigure[``Live'', variable-size, variable-position]{\fbox{\includegraphics[width=0.465\textwidth]{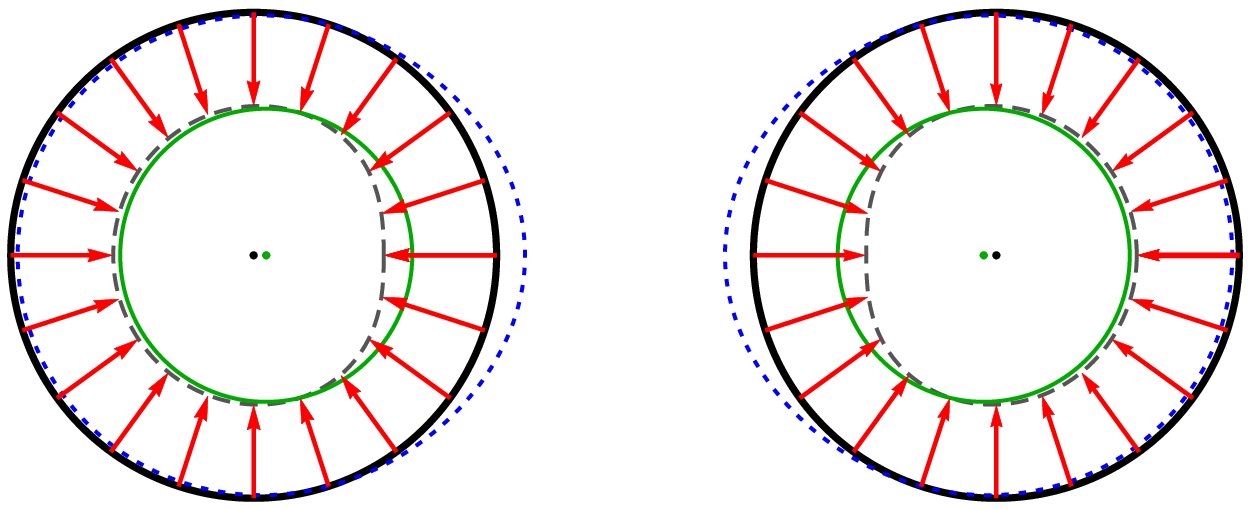}\label{twoc}}}
\caption{Two contractile spherical force dipoles in an infinite elastic medium: (a) ``Dead'' force dipoles applying an isotropic elastic force, (b-c) ``Live'' force dipoles regulating the force they apply in order to remain spherical even in the presence of the other force dipole. 
Red arrows are forces applied by each force dipole, black dashed lines are the corresponding self displacements, blue dotted lines are the displacements caused by the other force dipole, and green solid line is the total displacement. For illustration purposes, the initial distance between the spheres was set to $d=3 R_0$, and the self-displacement to $u_0=0.4 R_0$. The Poisson ratio is $\nu$=0.05. We show two of the ``live'' cases that we describe below, which differ in whether the volumes and positions of the spheres are preserved during the interaction (b) or not (c).}
\end{figure}

At first glance it seems that these results~\cite{whycellscare} contradict the previously-mentioned general result~\cite{02, 03} that bodies of any shape included in a continuous linear elastic material and applying on it isotropic dilational displacements on their surface \textit{do not interact}. In order to resolve this seeming contradiction, in this paper we compare the case of a spherical force dipole inside a concentric spherical Wigner-Seitz unit cell (see Fig.~\ref{WZ}) to the realistic configuration in which two spherical force dipoles are embedded at some distance $d$ between them in an infinite ECM with linear properties (see Fig.~\ref{circles3}). This comparison enables us to demonstrate that the reason for the contradiction is the difference between the definitions of the geometries in the two cases; In the two-sphere case the force dipoles obtain a drop-like shape due to the interaction [see Fig.~\ref{twoa}], while a spherical force dipole inside a spherical Wigner-Seitz unit cell preserves its spherical shape even when it interacts with other force dipoles.
Since the displacements created by each force dipole distort its neighbor, shape preservation may be achieved in the two-force-dipole setup only by application of appropriate anisotropic displacements by each dipole on its surface [see Fig.~\ref{twob},\ref{twoc}], and these give rise to the interaction energy that we study in this paper.

\section{Methods}

\subsection{Theoretical framework}

We consider two identical spheres of radius $R_0$ surrounded by an infinite material with linear elastic, or Hookean behavior defined by bulk modulus $K$ and shear modulus $G$ and denote the distance between their centers by $d$. In addition to the isotropic displacement $u_0$, each sphere applies an anisotropic displacement, which is intended to cancel the anisotropic displacements on its surface that are caused by the other sphere. In order to simplify the calculations we choose a left-handed coordinate system for sphere $2$; namely the newly defined angle $\theta_2$ equals $\pi-\theta'_2$, see Fig.~\ref{circles3}.

\subsection{General displacements generated by spherical force dipoles}

The displacement field around each force dipole must satisfy mechanical equilibrium, which we write in terms of the displacements field $\vec{u}$ as~\cite{01}:
\begin{equation}\label{Navier}
  \frac{1}{1-2 \nu} \nabla \nabla \cdot \overrightarrow{u} + \nabla^2 \overrightarrow{u} = 0.
\end{equation}
Due to the symmetry with respect to rotation about the axis connecting the centers of the two force dipoles, there is no dependence on the azimuthal angle $\phi$, thus we write Eq. (\ref{Navier}) in spherical coordinates as:
\begin{align}
\frac{1}{1-2 \nu} \frac{\partial}{\partial r} & \left[ \frac{1}{r^2} \frac{\partial}{\partial r} (r^2 u_r) + \frac{1}{r \sin{\theta} } \frac{\partial}{\partial \theta} (u_\theta \sin{\theta})\right] + \nabla^2 u_r - \frac{2}{r^2} u_r - \frac{2}{r^2} \frac{\partial{u_{\theta}}}{\partial{\theta}} - \frac{2 u_{\theta} \cot{\theta}}{r^2} = 0, \label{Navier21}\\
\frac{1}{1-2 \nu} \frac{1}{r} \frac{\partial}{\partial \theta} & \left[ \frac{1}{r^2} \frac{\partial}{\partial r} (r^2 u_r) + \frac{1}{r \sin{\theta} } \frac{\partial}{\partial \theta} (u_\theta \sin{\theta})\right] + \nabla^2 u_{\theta} + \frac{2}{r^2} \frac{\partial{u_{r}}}{\partial{\theta}}-\frac{u_{\theta} }{r^2 \sin^2 \theta}  = 0. \label{Navier22}
\end{align}
where the Laplacian in spherical coordinates excluding terms depending on $\phi$ is given by:
\begin{equation}\label{Laplacian}
  \nabla^2 = \frac{1}{r^2 \sin \theta} \left[ \frac{\partial}{\partial r} \left( r^2 \sin \theta  \frac{\partial}{\partial r} \right) + \frac{\partial}{\partial \theta} \left( \sin \theta \frac{\partial}{\partial \theta} \right) \right] .
\end{equation}

Based on the general solution for the displacement field of a sphere with given cylindrically symmetric displacements on its surface~\cite{01}, we write the anisotropic displacements field satisfying (\ref{Navier21}-\ref{Navier22}) outside the spherical force dipole ($r>R_0$) as a multipole expansion in terms of spherical harmonics $Y_n(\theta)=\sqrt{\frac{2n+1}{4 \pi}} P_n (\cos{\theta})$:
\begin{align}
  &u_{r i} = \frac{u_0 R_0^2}{r_i^2} +  u_0 \sum_{n=0}^{\infty} \left[ n (n+3-4 \nu) \frac{C_n R_0^n}{r_i^{n}} -  (n+1) \frac{D_n R_0^{n+2}} {r_i^{n+2}} \right] Y_n(\theta_i) , \label{uri} \\
  &u_{\theta i} =  u_0 \sum_{n=0}^{\infty} \left[  (-n+4-4 \nu) \frac{C_n R_0^n }{r_i^{n}} + \frac{D_n R_0^{n+2}} {r_i^{n+2}} \right] \frac{dY_n(\theta_i)}{d \theta_i} , \label{uti}
\end{align}
with $P_n(x)= 2^n \cdot \sum_{k=0}^{n} x^k \left(\begin{array}{c} n \\ k \\ \end{array}\right)\left(\begin{array}{c} \frac{n+k-1}{2} \\ n \\ \end{array} \right)$ the Legendre polynomial of order $n$~\cite{mathmetods}. Here $u_{r i}$ and $u_{\theta i}$ are the radial and angular components of the displacement field caused by sphere $i$, and the infinite sums represent the anisotropic corrections that each force dipole produces in order to cancel the shape distortion caused by its neighbor. $u_0$ and $R_0$ have been inserted so that the coefficients $C_n$ and $D_n$ are dimensionless.

Defining the dimensionless displacements $\widetilde{u}_{ri}=\frac{u_{ri}}{u_0}$, $\widetilde{u}_{\theta i}=\frac{u_{\theta i}}{u_0}$ and position $\widetilde{r}=\frac{r}{R_0}$, we rewrite Eqs. (\ref{uri}) and (\ref{uti}) in dimensionless form as:
\begin{align} \label{sphericalharmonicsnormalilzed1}
  &\widetilde{u}_{r i} (\tilde{r}_i,\theta_i) = \frac{1}{\widetilde{r}_i^2} + \sum_{n=0}^{\infty} \left[n (n+3-4 \nu) \frac{C_n}{\widetilde{r}_i^{n}} - (n+1) \frac{D_n} {\widetilde{r}_i^{n+2}} \right]  Y_n(\theta_i),  \\
  &\widetilde{u}_{\theta i} (\tilde{r}_i,\theta_i) =  \sum_{n=0}^{\infty} \left[ (-n+4-4 \nu) \frac{C_n}{\widetilde{r}_i^{n}}  + \frac{D_n} {\widetilde{r}_i^{n+2}} \right] \frac{dY_n(\theta_i)}{d \theta} \label{sphericalharmonicsnormalilzed2}
.\end{align}

 Note that (\ref{sphericalharmonicsnormalilzed1}-\ref{sphericalharmonicsnormalilzed2}) solve Eq. (\ref{Navier}) only when each force dipole is surrounded by an infinite homogeneous linearly-elastic medium, including in the interior of the neighboring force dipole. Biological cells clearly have a rigidity which differs from that of the ECM surrounding them, and thus this assumption seems to be problematic. We overcome this by realizing that we may first solve the mechanical problem in which the cells are assumed to have the same linear elastic properties as the ECM. The resultant solution includes a certain stress and displacement on the surface of each cell, and the solution outside the cells is independent on how the cell generates this stress on its surface. In particular, the stress that actual cells apply on their surrounding includes a passive stress coming from the rigidity of the cell plus an active stress coming from the external forces generated by molecular motors inside the cell. \emph{In our analysis we consider only the total stress and the work performed by it, which determines the interaction energy, and our results are valid irrespective of the mechanical rigidity of the cells themselves.}

\subsection{Cancellation condition}

For our ``live'' force dipoles the sum of the anisotropic displacements caused by the neighbor force dipole and of all the corrections applied by the discussed force dipole must vanish on its surface. From this we derive conditions for the coefficients $C_n$ and $D_n$ so that each force dipole will preserve its spherical shape even in the presence of the interaction with its neighboring force dipole. In order to apply the cancellation condition and to derive from it the expressions for $C_n$ and $D_n$, we transform the expressions for the displacement field of force dipole $2$ to the coordinate system of force dipole $1$ by substitution of the expressions for $r_2$ and $\theta_2$ in terms of $r_1$ and $\theta_1$ and then multiplying the displacement vector $\overrightarrow{u_2}=\left(u_{r2},u_{\theta 2}\right)$ by the rotation matrix:
\begin{equation}
\textbf{B}=
\left(
\begin{array}{cc}
 -\cos \left(\theta _1+\theta _2\right) & \sin
   \left(\theta _1+\theta _2\right) \\
 \sin \left(\theta _1+\theta _2\right) & \cos
   \left(\theta _1+\theta _2\right) \\
\end{array}
\right).
\end{equation}
We then write the resultant expressions for the radial and angular displacements caused by force dipole $2$ on the surface of force dipole $1$ in terms of the spherical harmonics of sphere $1$ by writing:
\begin{align}
  \left(u_{r} \right)_n &= \frac{2n+1}{2} \int_0^{\pi} u_r(\theta) Y_n(\theta)sin{\theta}d\theta,  \label{spheremem1}\\
  \left( u_{\theta} \right)_n &= \frac{2n+1}{2n(n+1)} \int_0^{\pi} u_{\theta}(\theta) \frac{Y_n(\theta)}{d\theta}sin{\theta}d\theta. \label{spheremem2}
\end{align}
As may be seen from (\ref{spheremem1},\ref{spheremem2}), every spherical-harmonic mode of force dipole $2$ contributes to all the modes on the surface of force dipole $1$.

As explained above, the sum of the anisotropic displacements caused by force dipole $2$ must be canceled on the surface of force dipole $1$ by the corrections that it applies. We write the dimensionless displacement $\widetilde{u}_{11}$ \textit{created by force dipole 1 on its surface } (namely at $\widetilde{r}_1=1$):
\begin{align}
 & \widetilde{u}_{r11}(\theta_1) \equiv \widetilde{u}_{r 1} (1,\theta_1) = \sum_{n=0}^{\infty} \left[n (n+3-4 \nu) C_n - (n+1) (D_n - \sqrt{4 \pi} \delta_{n,0})\right]  Y_n(\theta_1) , \label{surfr1} \\
 &\widetilde{u}_{\theta 11}(\theta_1) \equiv \widetilde{u}_{\theta 1} (1,\theta_1) =   \sum_{n=0}^{\infty}\left[ (-n+4-4 \nu) C_n + D_n \right] \frac{dY_n(\theta_i)}{d \theta} . \label{surft1}
\end{align}
The term $\delta_{n,0}$ in (\ref{surfr1}) is a Kronecker delta, which represents the isotropic radial displacement created by sphere 1 on its surface without the anisotropic cancellation corrections. This constant term does not depend on changes in the environment of the force dipole. The remaining terms are different modes of additional displacement created by this ``live'' force dipole in response to the displacement field induced on its surface by the neighboring force dipole. The dimensionless displacement $\widetilde{u}_{21}$ \textit{created by force dipole $2$ on the surface of force dipole $1$} is:
\begin{align}
  &\widetilde{u}_{r 21} (\theta_1)= \sum_{n=0}^{\infty} \sum_{m=0}^{\infty} \left[ f_{nm}^{Cr} (\widetilde{d}) C_m +  f_{nm}^{Dr} (\widetilde{d}) (D_m-\sqrt{4 \pi}\delta_{m,0}) \right] Y_n(\theta_1) \label{surfr2} , \\
  &\widetilde{u}_{\theta 21}(\theta_1) = \sum_{n=0}^{\infty} \sum_{m=0}^{\infty} \left[ f_{nm}^{C\theta} (\widetilde{d}) C_m +  f_{nm}^{D\theta} (\widetilde{d}) (D_m-\sqrt{4 \pi}\delta_{m,0}) \right] \frac{dY_n(\theta_1)}{d\theta_1}\label{surft2} ,
\end{align}
where the sum over $m$ originates from the fact that the displacement $\widetilde{u}_2$ created by force dipole 2 is given by a multipole expansion (\ref{sphericalharmonicsnormalilzed1}-\ref{sphericalharmonicsnormalilzed2}) with the corrective magnitudes $C_m$ and $D_m$. The sum over $n$ originates from the fact that after the coordinate transformation, when these modes are expressed in terms of the spherical harmonics in the coordinate system of force dipole 1, each mode from force dipole 2 contributes to all the modes of force dipole 1. The functions $f_{nm}^{Cr}(\widetilde{d})$, $f_{nm}^{Dr}(\widetilde{d})$, $f_{nm}^{C\theta}(\widetilde{d})$ and $f_{nm}^{D\theta}(\widetilde{d})$, given in Appendix~\ref{appb} depend only on the dimensionless distance $\widetilde{d}=\frac{d}{R_0}$ between the spherical force dipoles.

We now require that for ``live'' force dipoles the total displacement $\widetilde{u}_{11}(\theta_1)+\widetilde{u}_{21}(\theta_1)$ on the surface of force dipole 1 is isotropic. We begin by considering the simplest (but strictest) regulation scenario, for which not only is this total displacement isotropic, but its magnitude remains equal to the displacement $u_0$ in the absence of interactions between the force dipoles. Moreover, we require that the center of symmetry of each force dipole does not move. We will later consider three additional scenarios in which the interaction causes the force dipoles to change their volume and/or to move, yet they remain spherically symmetric. Thus at this point we require that:
\begin{align}
  \widetilde{u}_{r11}(\theta_1)+\widetilde{u}_{r21}(\theta_1)& \equiv 1 , \label{gencancel1}\\
  \widetilde{u}_{\theta11}(\theta_1)+\widetilde{u}_{\theta21}(\theta_1)& \equiv 0 . \label{gencancel2}
\end{align}
Substituting (\ref{surfr1},\ref{surft1},\ref{surfr2},\ref{surft2}) in (\ref{gencancel1},\ref{gencancel2}) yields:
\begin{align}
&\begin{gathered}\label{fincancel1}
  \sum_{n=0}^{\infty}  \left\{\left[n (n+3-4 \nu) C_n - (n+1) (D_n - \sqrt{4 \pi} \delta_{n,0})\right] +  \sum_{m=0}^{\infty} \left[ f_{nm}^{Cr} (\widetilde{d}) C_m  + f_{nm}^{Dr} (\widetilde{d}) (D_m-\sqrt{4 \pi}\delta_{m,0}) \right] \right\} Y_n(\theta_1) = 1 ,
\end{gathered}\\
& \begin{gathered}\label{fincancel2}
  \sum_{n=0}^{\infty}\left\{ \left[ (-n+4-4 \nu) C_n + D_n \right] + \sum_{m=0}^{\infty} \left[ f_{nm}^{C\theta} (\widetilde{d}) C_m + f_{nm}^{D\theta} (\widetilde{d}) (D_m-\sqrt{4 \pi}\delta_{m,0}) \right] \right\} \frac{dY_n(\theta_1)}{d\theta_1} = 0 .
\end{gathered}
\end{align}
Due to the orthogonality of the Legendre polynomials, for these infinite sums to satisfy these cancellation conditions, each term in the sums must cancel independently. Thus for all $n \ge 1$ we require:
\begin{align}
  &n (n+3-4 \nu) C_n - (n+1) D_n + \sum_{m=0}^{\infty} \left[ f_{nm}^{Cr} C_m + f_{nm}^{Dr} (D_m - \sqrt{4 \pi} \delta_{m,0}) \right] = 0, \label{surf1} \\
  &(-n+4-4 \nu) C_n + D_n + \sum_{m=0}^{\infty} \left[ f_{nm}^{C\theta} C_m + f_{nm}^{D\theta} (D_m - \sqrt{4 \pi} \delta_{m,0}) \right] = 0. \label{surf2}
\end{align}
 Note that for $n=0$, from (\ref{sphericalharmonicsnormalilzed1}-\ref{sphericalharmonicsnormalilzed2}) $C_0$ is irrelevant, thus we set it to zero. Moreover, since $Y_0(\theta_1)=1$, $\frac{dY_0(\theta_1)}{d\theta_1}=0$ and (\ref{fincancel2}) holds trivially, thus for $n=0$ we obtain only one equation, from Eq. (\ref{fincancel1}):
\begin{align}\label{0eq}
-D_0 + \sum_{m=0}^{\infty} \left[ f_{0m}^{Cr} C_m + f_{0m}^{Dr} (D_m - \sqrt{4 \pi} \delta_{m,0}) \right] = 1.
\end{align}

We obtain closure of the infinite coupled linear Eqs.~(\ref{surf1}, \ref{surf2}, \ref{0eq}) by assuming that $C_n=0$ and $D_n=0$ for $n>n_{max}$. This is justified since we will be interested in large separations between the force dipoles, and due to the fact that the solutions decay as $1/r^{n}$, thus at large $r$, large $n$ terms become negligible. We verify this numerically by increasing $n_{max}$ until convergence, see Section~\ref{results} below.

\subsection{The n=0,1 modes}\label{n01}

\begin{figure}[h]
\centering
\includegraphics[trim=0mm 65mm 0mm 35mm,width=0.6\columnwidth]{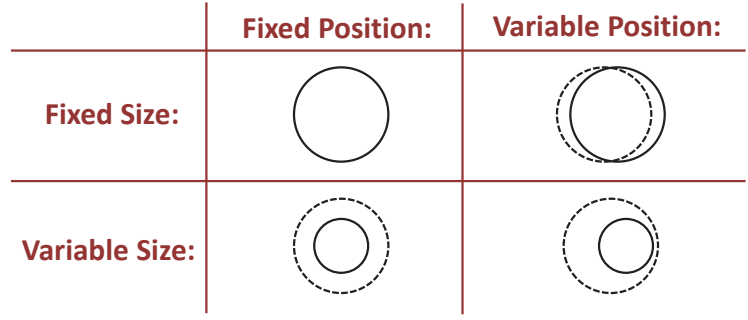}
\caption{The four possible scenarios of shape regulation, depending on whether the size ($n=0$) and the position ($n=1$) are regulated or not.}\label{scenarios_diagram}
\end{figure}

Each of the spherical harmonics $Y_n(\theta)$ represents a different mode of deformation of the force dipole. For scalar fields the first two modes represent volume change ($n=0$) and translation of the center of symmetry ($n=1$). However, as can be seen from Eqs. (\ref{sphericalharmonicsnormalilzed1},\ref{sphericalharmonicsnormalilzed2}), since the displacement field is vectorial, here the $n=1$ mode may represent a combination of translation and deformation. Since our definition of ``live'' force dipoles focuses only on shape regulation, inclusion in the solution of the terms $n=0$ and the part of the $n=1$ term that controls translation is not mandatory. Inclusion of one or both of these terms will result in complementary condition(s) in addition to the shape preservation condition and will describe a different type of self-regulation of the force dipoles. Overall there are four combinations for the inclusion or exclusion of the $n=0$ and $n=1$ terms and thus we analyze four possible scenarios of homeostasis or self-regulation, see Fig.~\ref{scenarios_diagram} and Fig.~\ref{twob}-\ref{twoc} above.

As a result of exclusion of one or more of the terms of the sum, Eqs.~(\ref{surf1}-\ref{0eq}) will not vanish anymore at the corresponding modes, i.e. non-zero resultant displacement will exist. At the same time the force applied by the dipole and the resulting elastic stress will vanish at these modes. Mode $n=0$ of the displacement never vanishes since it includes the constant term $u_0$ which represents the fixed symmetric displacement created by the force dipole, which does not depend on the interaction. For the size $(n=0)$ we have two options - fixed size (FS) namely displacement homeostasis, in which we regulate this mode, and variable size (VS) namely stress homeostasis, in which we allow it to have an additional contribution due to the interaction. Similarly we consider either fixed position (FP) or variable position (VP) of the force dipole $(n=1)$. In the VP case instead of the cancellation condition (\ref{surf1}-\ref{surf2}), the coefficients $C_1$, $D_1$ obey the following:
\begin{align}
  &(4-4 \nu) C_1 - 2 D_1 + \sum_{m=0}^{\infty} ( f_{1m}^{Cr} C_m + f_{1m}^{Dr} D_m ) = \sum_{m=0}^{\infty} \left( \frac{f_{1m}^{Cr}+ f_{1m}^{C\theta}}{2} C_m + \frac{f_{1m}^{Dr} + f_{1m}^{D\theta}}{2} D_m \right), \label{c1d11} \\
  &(3-4 \nu) C_1 + D_1 + \sum_{m=0}^{\infty} ( f_{1m}^{C\theta} C_m + f_{1m}^{D\theta} D_m) = \sum_{m=0}^{\infty} \left( \frac{f_{1m}^{Cr}+ f_{1m}^{C\theta}}{2} C_m + \frac{f_{1m}^{Dr} + f_{1m}^{D\theta}}{2} D_m \right). \label{c1d12}
\end{align}

Equations (\ref{c1d11}-\ref{c1d12}) allow us to find $C_1$ and $D_1$ that cancel the deformation at mode $1$ without canceling the translational motion of the force dipole. They follow from Eqs. (\ref{surf1}-\ref{surf2}) and from the fact that for translation the coefficients of $P_1(cos\theta)=cos(\theta)$ in the radial direction and of $\frac{dP_1(cos\theta)}{d\theta} =sin(\theta)$ in the tangential direction must be equal.

\subsection{Interaction energy}\label{Eoie}

After obtaining the coefficients $C_n$ and $D_n$ and substituting them in Eqs.~(\ref{surfr1}-\ref{surft2}) we compute the interaction energy from the work performed by the force dipoles to generate this deformation:
\begin{align}\label{inten}
   \Delta E &=E-2E_0 =\frac{1}{2} \int_S \left( \overrightarrow{u} \cdot \overrightarrow{F} - \overrightarrow{u_0} \cdot \overrightarrow{F_0} \right) d s.
\end{align}
Here the integration is over the surfaces of both spheres, $\overrightarrow{F}$ is the force per unit area applied by each force dipole on its environment.
It is important to emphasize the difference in calculation of the interaction energies in the cases of dead versus live force dipoles. In the general case of two dead force dipoles the displacements and forces applied by each one of them are not affected by changes in its environment and thus the amount of additional work done by it equals~\cite{Eshelby87} [see Eq.~(\ref{inten})]:
\begin{equation}\label{Eshelbyint}
  \Delta E_i=\frac{1}{2}\int_{S_i} \left[ (\overrightarrow{u_i}+\overrightarrow{u_j}) \cdot \overrightarrow{F_i}-\overrightarrow{u_i} \cdot \overrightarrow{F_i} \right] dS_i = \frac{1}{2}\int_{S_i}  \overrightarrow{u_j} \cdot \overrightarrow{F_i} dS_i .
\end{equation}
In the particular case that the force dipoles apply isotropic forces or displacements causing only volume change without distortion (as for two ``dead'' spherical force dipoles) the interaction energy $\Delta E$ vanishes~\cite{02}. In contrast, for two ``live'' force dipoles the forces and displacements created by each one of them are modified due to the interaction with the neighbor. The resultant forces and displacements applied by ``live'' force dipoles are not isotropic and the interaction energy does not vanish anymore. Here the additional work that is performed by each force dipole is:
\begin{align}\label{Ourint}
  \Delta E_i &=\frac{1}{2}\int_{S_i} \left[ (\overrightarrow{u_i}+\delta \overrightarrow{u_i}+\overrightarrow{u_j}+\delta \overrightarrow{u_j}) (\overrightarrow{F_i}+\delta \overrightarrow{F_i}) - \overrightarrow{u_i} \cdot \overrightarrow{F_i} \right] dS_i  \nonumber \\
  &=\frac{1}{2}\int_{S_i} \left[ \overrightarrow{u_j} \cdot \overrightarrow{F_i} + \delta \overrightarrow{u_i} \cdot \overrightarrow{F_i} +\delta \overrightarrow{u_j} \cdot \overrightarrow{F_i} + \overrightarrow{u_i} \cdot \delta \overrightarrow{F_i} + \overrightarrow{u_j} \cdot \delta \overrightarrow{F_i} + \delta \overrightarrow{u_i} \cdot \delta \overrightarrow{F_i}  + \delta \overrightarrow{u_j} \cdot \delta \overrightarrow{F_i} \right] dS_i .
\end{align}
Here $\delta \overrightarrow{u}$ and $\delta \overrightarrow{F}$ are the ``live'' parts of the displacement and the force created by the force dipoles due to their interaction. For two ``live'' force dipoles, we evaluate the interaction energy using the stress tensor $\underline{\underline{\tau}}$ that arises in the elastic environment of each force dipole in response to the displacement $\overrightarrow{u}$ on its surface. The energy evaluated in Eq. (\ref{Ourint}) equals the work done by spherical force dipole $i$ on its surface and thus we add the index $i$ and the total interaction energy is $E_1+E_2$. The forces applied by each force dipole are equal and opposite to the forces applied on it by the environment: $\overrightarrow{F}= - \underline{\underline{\tau}} \cdot \hat{r}$\label{force}, where $\hat{r}$ is the outward pointing unit vector normal to the surface of the sphere after its deformation and movement. Since we allow only volume change and translational motion of the entire force dipoles, the direction of the normal does not change due to the interaction. The terms of the stress tensor $\underline{\underline{\tau}}$ based on the general solution of the displacements field Eqs.~(\ref{uri}-\ref{uti})~\cite{01} are given in Appendix~\ref{appc}. $\overrightarrow{F_0}=\frac{4Gu_0}{R_0}\hat{r}$ is the force per unit area on the surface of the sphere in the case of a single force dipole with known isotropic displacement $\overrightarrow{u_0} = u_0 \hat{r}$ on its surface and without interactions with other spheres, and thus the self energy is $E_0 = 8 \pi G u_0^2 R_0$\label{E0}.

Evaluation of the integral in Eq.~(\ref{inten}) may be simplified by noting that both the displacements $\overrightarrow{u}$ and the stresses $\underline{\underline{\tau}}$ are expressed in terms of Legendre polynomials and their derivatives [see Eqs. (\ref{surf1}-\ref{surf2})] and using their orthogonality we get (see Appendix~\ref{appe}):
\begin{align}\label{Eint}
  E = 2 \cdot \frac{1}{2}\int_S \overrightarrow{u} \cdot \overrightarrow{F} ds = 2 E_0 \left( 1 + D_0 - \left\{ - D_0 + \sum_{m=0}^{\infty} \left[ f_{0m}^{Cr} C_m + f_{0m}^{Dr} (D_m - \sqrt{4\pi} \delta_{m,0}) \right] \right\} \right) .
\end{align}
In the FS cases $D_0$ is evaluated from the cancellation condition and Eq.~(\ref{Eint}) becomes:
\begin{equation}\label{ufsa}
  E_{FS} = 2 E_0 \left( 1 + D_0 \right) = 2 E_0 \left( 1 - \left\{ \sum_{m=0}^{\infty} \left[ f_{0m}^{Cr} C_m + f_{0m}^{Dr} \left(D_m - \sqrt{4\pi}\delta_{m,0}\right) \right] \right\} \right) ,
\end{equation}
while in the VS cases $D_0$ is zero and Eq.~(\ref{Eint}) becomes:
\begin{equation}\label{ufsb}
  E_{VS} = 2 E_0 \left\{ 1 + \sum_{m=0}^{\infty} \left[ f_{0m}^{Cr} C_m + f_{0m}^{Dr} \left(D_m - \sqrt{4\pi} \delta_{m,0}\right) \right] \right\} .
\end{equation}
The meaning of the sum of the last two terms in Eqs. (\ref{ufsa}-\ref{ufsb}) is the change of the volume of force dipole 1 caused by all the modes of displacement created by force dipole 2 (see Fig. \ref{circles3}). We thus define:
\begin{equation}\label{dV21}
  \Delta V_{21} \equiv \sum_{m=0}^{\infty} \left[ f_{0m}^{Cr} C_m + f_{0m}^{Dr} \left(D_m - \sqrt{4\pi}\delta_{m,0}\right) \right] .
\end{equation}
Thus the interaction energies in the FS cases become:
\begin{equation}\label{DeltaE1}
  \Delta E_{FS} =- 2 E_0 \Delta V_{21} ,
\end{equation}
and in the VS cases:
\begin{equation}\label{DeltaE2}
  \Delta E_{VS} = 2 E_0 \Delta V_{21} .
\end{equation}

\section{Results}\label{results}

\begin{figure}[b]
\centering
\subfigure{\includegraphics[trim= 25mm 150mm 25mm 50mm,width=0.49\textwidth]{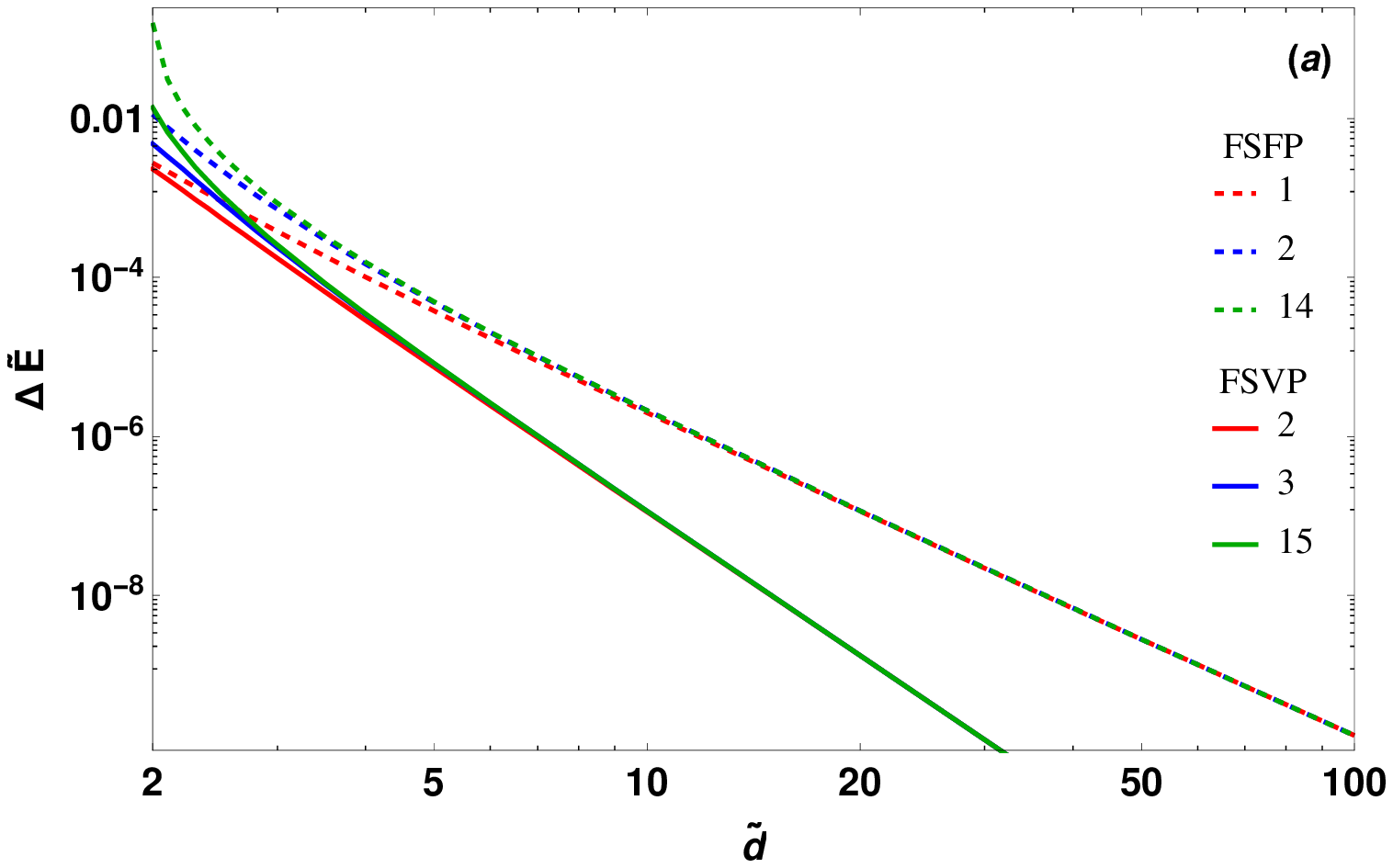}\label{figa}}
\subfigure{\includegraphics[trim= 25mm 150mm 25mm 50mm,width=0.49\textwidth]{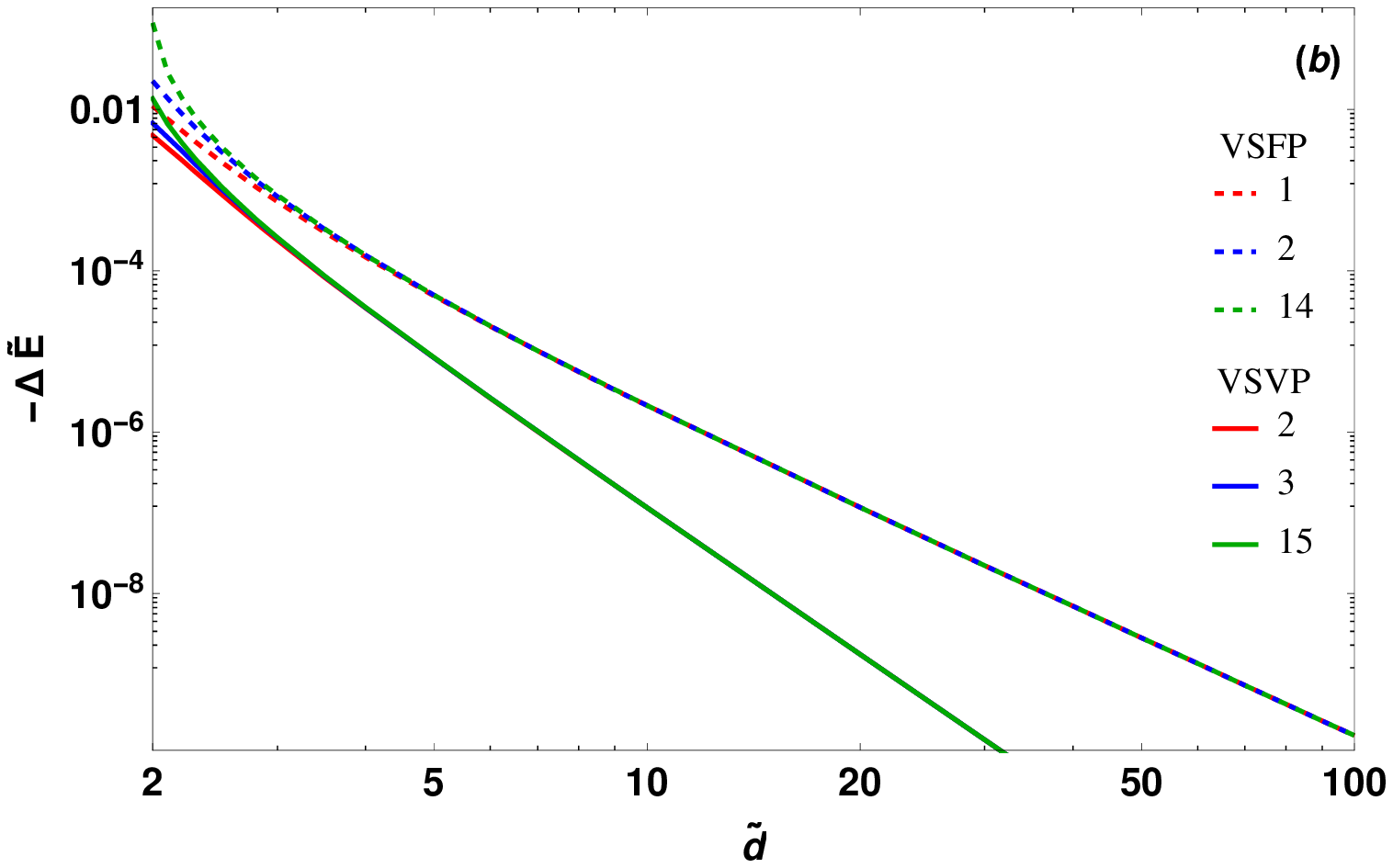}\label{figb}}
\caption{Dimensionless interaction energy $\Delta \widetilde{E}=\frac{\Delta E}{E_0}$ vs dimensionless distance between the two force dipoles $\widetilde{d}=\frac{d}{R_0}$, with different values of $n_{max}$ as indicated in the legend. Poisson ratio was set to $\nu=0.45$. Note that the interaction energy in the FS cases is positive (repulsive), and in the VS cases negative (attractive). The far-field behavior in the FP cases is given by $1/\tilde{d}^4$ and for VP by $1/\tilde{d}^6$.}\label{alphabeta}
\end{figure}

\begin{figure}[t]
\includegraphics[width=0.55\textwidth] {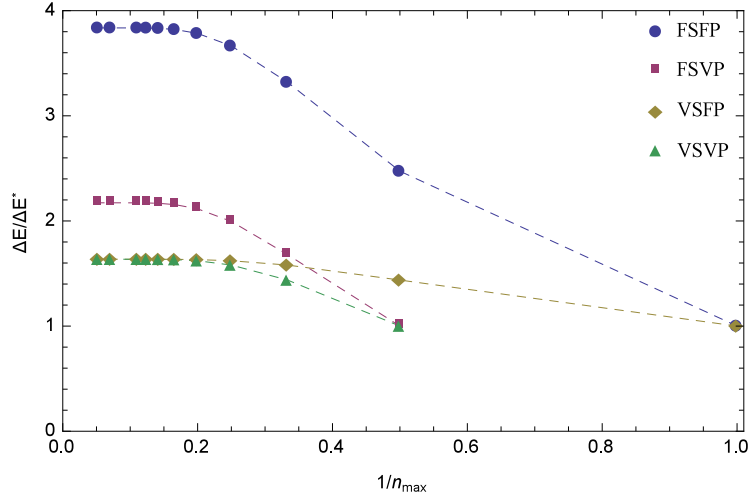}
\caption{Interaction energy vs inverse of highest term $n_{max}$ included in the solution. Energy is rescaled by the result with the minimal possible value of $n_{max}$. Results are shown for Poisson ratio $\nu=0.45$ and dimensionless distance $\tilde{d}=2.5$ between the centers of the two force dipoles.}\label{fig:convergence}
\end{figure}

We evaluated the interaction energy by terminating the infinite sums (\ref{surf1}-\ref{surf2}) at different values of $n_{max}$, up to $n_{max}=15$ for all four regulation scenarios. Figure~\ref{alphabeta} shows the normalized interaction energy $\Delta\tilde{E} = \Delta E / E_0$ vs the normalized distance between the force dipoles $\tilde{d} = d/R_0$ for all four regulation scenarios. Figure~\ref{fig:convergence} shows the convergence of the interaction energy with increasing the value of $n_{max}$. Even at the smallest dipole-dipole separations the solution converges with moderate value of $n_{max}=4-5$. Moreover, for distances larger than 4 sphere radii in the FSVP, VSFP and VSVP cases and for distances larger than 6 sphere radii in the FSFP case the energy is well approximated by taking the minimal $n_{max}$ possible. Namely for the FP cases $n_{max}=1$ suffices, and for the VP cases we take $n_{max}=2$, see Table~\ref{tabcoef}. Since in Eq.~(\ref{surf1}) the coefficient of $C_0$ vanishes and Eq.~(\ref{surf2}) vanishes for $n=0$ there is no need to evaluate the coefficient $C_0$. Thus in the FS cases we get three coupled linear equations and in the VS cases two coupled linear equations. It is important to reiterate the role of the $n=0$ and $n=1$ terms in the solution. Excluding the $n=1$ term of the sum in Eqs.~(\ref{uri}-\ref{uti}) constitutes the VP cases in which the force dipoles are free to move but preserve their size and shape, and excluding the $n=0$ term constitutes the VS cases in which the force dipoles preserve their spherical shape but are free to change their volume due to the interaction. We see from Eq.~(\ref{Eint}) that the presence of the coefficient $D_0$ determines the sign of the interaction energy, see also Fig.~\ref{alphabeta} and Table~\ref{tabcoef}; \emph{Force dipoles with FS regulation ($D_0 \ne 0$), are repelled, while force dipoles with VS regulation ($D_0 = 0$) are attracted.}

\begin{table}[t]
\begin{center}
\begin{tabular}{| c | c | c | c | c | c | c |}
    \hline
      & Fixed Size & Fixed Size & Variable Size & Variable Size \\
      & Fixed Position & Variable Position & Fixed Position & Variable Position \\
      & ($FSFP$) & ($FSVP$) & ($VSFP$) & ($VSVP$) \\  \hline
      $n_{max}$ & 1 & 2 & 1 & 2 \\ \hline
    $D_0$ & $\frac{(1-2 \nu) }{5-6\nu}\frac{1}{\widetilde{d}^4}$ & $\frac{5(1-2\nu)}{4-5\nu}\frac{1}{\widetilde{d}^6}$ & 0 & 0 \\[1.3ex] \hline
    $C_1$ & $-\frac{3}{2 (5-6\nu)}\frac{1}{\widetilde{d}^2}$ & $\frac{9 (2-3\nu)}{4(4-5\nu)(5-6\nu)}\frac{1}{\widetilde{d}^7}$ & $-\frac{3}{2 (5-6\nu)}\frac{1}{\widetilde{d}^2    }$ & $\frac{9 (2-3\nu)}{4(4-5\nu)(5-6\nu)}\frac{1}{\widetilde{d}^7}$\\[1.3ex] \hline
    $D_1$ & $-\frac{1}{2 (5-6\nu)}\frac{1}{\widetilde{d}^2}$ & $\frac{9 (2-3\nu)(7-8\nu)}{4(4-5\nu)(5-6\nu)}\frac{1}{\widetilde{d}^7}$ & $-\frac{1}{2 (5-6\nu)}\frac{1}{\widetilde{d}^2}$ & $\frac{9 (2-3\nu)(7-8\nu)}{4(4-5\nu)(5-6\nu)}\frac{1}{\widetilde{d}^7}$ \\ [1.3ex]\hline
    $C_2$ & 0 & $-\frac{5}{4(4-5\nu)}\frac{1}{\widetilde{d}^3}$ & 0 & $-\frac{5}{4(4-5\nu)}\frac{1}{\widetilde{d}^3}$ \\[1.3ex] \hline
    $D_2$ & 0 & $-\frac{3}{2(4-5\nu)}\frac{1}{\widetilde{d}^3}$ & 0 & $-\frac{3}{2(4-5\nu)}\frac{1}{\widetilde{d}^3}$ \\[1.3ex] \hline
    $\Delta V_{21}$ & $-\frac{2 (1 - 2 \nu) C_1}{3 \widetilde{d}^2}$ & $-\frac{4(1-2\nu) C_2}{\widetilde{d}^3}$ & $-\frac{2 (1 - 2 \nu) C_1}{3 \widetilde{d}^2}$ & $-\frac{4(1-2\nu) C_2}{\widetilde{d}^3}$ \\ \hline
    $\Delta \widetilde{E}_{\infty}$ & $\frac{2 (1-2 \nu)
}{(5-6\nu)}\cdot \frac{1}{\widetilde{d}^{4}}$ & $\frac{10 (1-2 \nu)}{(4-5\nu)} \cdot \frac{1}{\widetilde{d}^{6}}$ & $-\frac{2 (1-2 \nu)
}{(5-6\nu)} \cdot \frac{1}{\widetilde{d}^{4}}$ & $-\frac{10 (1-2 \nu) }{(4-5\nu)}\cdot \frac{1}{\widetilde{d}^{6}}$ \\ \hline
\end{tabular}
\caption{Coefficients $C_n$ and $D_n$ of the multipole expansion at asymptotically long-distances, $\widetilde{d} \gg 1$. Here $n_{max}$ defines the highest order term taken into account, and $\Delta \widetilde{E}_{\infty}$ is the asymptotic long-distance behavior of the resultant dimensionless interaction energy $\Delta \widetilde{E} \equiv \frac{E_{int}}{E_0}$. $\widetilde{d}=\frac{d}{R_0}$ is the dimensionless distance between the force dipoles, and $\Delta V_{21}$ is defined in Eq.~(\ref{DeltaE2}). The corresponding expressions for arbitrary distance are given in Appendix~\ref{appf}.} \label{tabcoef}
\end{center}
\end{table}

This is similar to our analysis within the spherical unit-cell approximation of different homeostasis scenarios, where we found repulsion when the force dipole generates a fixed displacement on its boundary and attraction when it applies a fixed stress on its environment~\cite{whycellscare}. Relating this to our present results, in the FS cases the resultant displacement on the boundary of each force dipole is fixed and does not depend on the changes in its environment, which corresponds to displacement homeostasis. In the VS cases the size of each force dipole changes due to the interaction, namely the displacement is not fixed. There is no correction to the $n=0$ mode, which means that there is no additional active force or stress at this mode, and we relate this to stress homeostasis, in which indeed the interaction is attractive.

An intuitive explanation for the sign difference in the interaction may be considered as follows; FS means that in the proximity to other force dipoles, each force dipole has to exert an additional force in order to generate the displacement that it was programmed to have. Thus more energy is required and the force dipoles would benefit energetically from moving away from each other. As noted above, VS is related to stress homeostasis, for which the tension that neighboring contractile force dipoles generate around themselves add up, and less energy is required in order to reach the homeostatic stress value, thus it is beneficial for cells to be close to other cells.

\emph{The position regulation $(n=1)$ sets the strength of these attractive or repulsive interactions}. For VP, $C_1=D_1=0$ and at long distances the interaction energy decays as $1/d^6$, while for FP, $C_1$ and $D_1$ are nonzero, which leads to a $1/d^4$ decay, see Fig.~\ref{alphabeta} and Table~\ref{tabcoef}. To quantitatively relate these functional forms to the results of our spherical unit-cell approximation, we consider an infinite array of active force dipoles. Using our results for the interaction between two active force dipoles, we now evaluate the interaction energy in an infinite 3D array of such shape-regulating spherical force dipoles (see Fig. \ref{WZ}). In a linear medium and for small displacements generated by each force dipole, by superposition the total interaction energy per force dipole is:
\begin{equation}\label{Etot2}
  E_{tot}=\int_{r=0}^{\infty} n E_{int}(r) \cdot 4\pi r^2 dr\approx\int_{r=D}^{\infty} n E_{int}(r) \cdot 4\pi r^2 dr ,
\end{equation}
where $n$ is the density of force dipoles at distance $r$ from any given force dipole, which for large distances we assume is uniform, and we introduce a near-field cutoff distance equal to the lattice spacing $D$.

Substituting the expressions for the two-force-dipole interactions from Table~\ref{tabcoef} in Eq.~(\ref{Etot2}) and performing the integrations, yields: $|E_{tot}| \propto 1/D$ (FP), $1/D^3$ (VP). The VP cases are quantitatively consistent with the results of our previous work within the spherical unit-cell approximation~\cite{whycellscare} that the interaction energy in an array of ``live'' force dipoles scales as $1/R_c^3$, since the radius $R_c$ of the periodic unit cell is related to the distance $D$ between force dipoles in the periodic array. Due to the symmetry of the spherical unit-cell setup, by definition there is no need for the force dipole to exert any force in order to remain in place and thus there is no regulation of the position, which corresponds to our VP case.

The expressions for the interaction energy given in Table \ref{tabcoef} (and similarly in Table~\ref{tabcoef2} in Appendix~\ref{appf} for arbitrary distance) all vanish in the incompressible limit ($\nu=0.5$). The self-displacement generated by each force dipole separately includes a pure compressive mode inside that force dipole, and sustaining this in an incompressible medium with a fixed displacement $u_0$ on the boundary requires an infinite force. Thus the problem of mechanical interactions between force dipoles in a completely incompressible medium deserves a separate analysis, which we defer to future work.

\section{Discussion}

We model live cells as spherical force dipoles surrounded by a linear, or Hookean elastic environment. Since in the case of isotropic displacements and forces the interaction energy between force dipoles vanishes, we distinguish between ``dead'' behavior in which the force dipoles apply constant forces and self-displacements on their surface, and ``live'' behavior in which the forces and self-displacements applied change in response to changes in the environment of the force dipoles. We solved the interaction energy for four different types of such regulation in which the force dipoles preserve their spherical shape and in addition volume, position, both or none of them. We found the interaction energy to be inversely proportional to the distance between the force dipoles to the fourth power in the case of fixed position and to the sixth power in the case of variable position. These results are similar to and stem from the same reasons as the van der Waals dipole-induced dipole interaction in electrostatics~\cite{Israelachvili}. We also found that in the case of volume preservation the force dipoles are repelled while without volume preservation they are attracted.

We solved the deformation fields for the case when the rigidity of the force dipoles is identical to that of their environment. However, biological cells are complex entities and their rigidity differs from point to point and also differs from the rigidity of the extracellular matrix. In order to relate our results to live cells we describe each of them as a mechanism that applies forces on its environment on the surface and responds by their variation to application of external force or displacement. The displacements and forces applied by a cell may be divided into ``passive'' and ``active'' parts. The ``passive'' part of the forces or displacements would stay the same if the cells were dead and preserve their elastic properties, while the ``active'' part depends on the programmed behavior of the cells and is generated by the contraction of acto-myosin networks inside them. Since the resultant force and displacement are the sum of those two parts cells may create such ``active'' response such that the resultant forces and displacements will coincide with the case considered here, for which their rigidity coincides with the rigidity of their environment.

Stress homeostasis vs displacement homeostasis is believed to occur in different cell types and in different mechanical environments~\cite{C5IB00013K}. We suggest to examine in experiments the correlation that we find here between these distinct regulatory behaviors and attraction vs repulsion. Specifically, for cells that are known to regulate their stress or displacement one could test whether they indeed are attracted or repelled, respectively. Alternatively, when the regulatory behavior is not known, we suggest that our results will enable to infer it from the direction of the interaction. It would be interesting to consider also more complicated regulatory behaviors on top of the two extreme limits of fixed stress and fixed displacement. In particular, it would be interesting to test whether our fixed-position vs variable-position cases could generate more complex scenarios, since we predict that that would relate to the interaction strength. Clearly the interaction energy that we focus our analysis on may not be directly probed in experiments. We suggest that our predictions on attraction or repulsion between cells could be tested by studying other experimental indications of inter-cellular interactions. For instance, cells that are attracted to each other tend to try and move closer together but in rigid environments may not move and instead send protrusions one toward each other~\cite{Reinhart-King2008}. Moreover, focusing of mechanical stress between interacting cells may be experimentally visualized~\cite{Winer2009}.

Following our work on three-dimensional spherical force dipoles, it would be interesting to solve the case of two-dimensional disks on an elastic medium. The same four regulatory possibilities can be considered. This could more directly be related to experiments in which cells are grown on surfaces of different elastic materials~\cite{Reinhart-King2008}. We expect that in this case the interaction energy will not vanish even for ``dead'' force dipoles. Clearly the work presented in this paper is a first step toward theoretical analysis of interactions that stem from mechanobiological regulation, and in particular shape regulation of live cells. Our focus on linearly-elastic, homogeneous and isotropic media enabled us to obtain the detailed analytical description presented above. It would be important to continue this line of research and study the influence of the material nonlinearity, heterogeneity and anisotropy of the ECM on the interaction between cells.

\acknowledgments

We thank Dan Ben-Yaakov, Kinjal Dasbiswas, Haim Diamant, Erez Kaufman, Ayelet Lesman, Sam Safran, Nimrod Segall, Eial Teomy, Daphne Weihs and Xinpeng Xu for helpful discussions. This work was partially supported by the Israel Science Foundation grant No. 968/16, by a grant from the United States-Israel Binational Science Foundation and by a grant from the Ela Kodesz Institute for Medical Physics and Engineering.

\appendix

\section{Vanishing interaction energy between ``dead'' cells}\label{appa}

Here we calculate the vanishing interaction energy between two ``dead'' spherical force dipoles each with radius $R_0$ applying an isotropic radial force $F$ on its environment, see Fig. \ref{circles3}. The environment behaves as a linear elastic material with shear modulus $G$. The displacements field around a single such force dipole is given by~\cite{Dau}:
\begin{equation}\label{u}
  \overrightarrow{u}(R)=\frac{F R_0^3}{4 G r^2} \widehat{r} .
\end{equation}
In order to evaluate the resultant displacement of two such force dipoles we superimpose the displacement fields of both force dipoles. To do so, we translate the displacement field created by force dipole 1 to the coordinate system of force dipole 2. Using the cosine theorem we rewrite the displacement fields created by the force dipoles in terms of $\theta_2$ and $d$, see Fig. \ref{circles3}:
\begin{eqnarray}
  \overrightarrow{u_1} &=& \frac{F R_0^3}{4 G \left( R_0^2+d^2+2R_0d \cos{\theta_2} \right)} \widehat{r_1} , \\
  \overrightarrow{u_2} &=& \frac{F R_0}{4 G} \widehat{r_2} .
\end{eqnarray}
Here $\widehat{r_1}$ and $\widehat{r_2}$ are unit vectors in the radial directions of force dipoles 1 and 2, respectively. The resultant displacement in the radial direction of force dipole 2 is:
\begin{eqnarray}
  u_r = \left[ \frac{F R}{4 G} \widehat{r_2} + \frac{F R_0^3}{4 G \left( R_0^2+d^2+2R_0d \cos{\theta_2} \right)} \widehat{r_1} \right] \cdot \widehat{r_2}  =  \frac{F R_0}{4 G} \left[1 + \frac{R_0^2 \left( R_0 + d \cos{\theta_2} \right)}{\left( R_0^2+d^2+2ad \cos{\theta_2} \right)^{3/2}} \right] .
\end{eqnarray}
The symmetry of the system imposes that the total work done by the system reads:
\begin{align}
  E = \int_0^{\pi} \left( 2 \cdot \frac{1}{2}  F u_r \cdot 2\pi R_0 \sin{\theta_2} \right) R_0 d \theta_2
  = \frac{\pi F^2 R_0^3}{2G} \int_0^{\pi} \left[\sin{\theta_2} + \frac{\left( R_0 + d \cos{\theta_2} \right) R_0^2 \sin{\theta_2}}{\left( R_0^2+d^2+2R_0d \cos{\theta_2} \right)^{3/2}} \right] d \theta_2 .
\end{align}
After evaluation we obtain:
\begin{equation}\label{W2W1}
  E = \frac{ \pi F^2 R_0^3}{2G}=2 \cdot  4 \pi R_0^2 \cdot \frac{1}{2} F \cdot \frac{F R_0}{4 G} = 2 E_0 .
\end{equation}
Thus the interaction energy vanishes in this case. A similar analysis may be done for two force dipoles which apply radial displacements $u_0$ on their surfaces. The energy in this case may be written as:
\begin{equation}\label{W2W2}
  E =16 R_0 \pi u_0^2 G = 2 E_0 ,
\end{equation}
and also here the interaction energy vanishes.

\section{The functions $f_{nm}^{Cr},\ f_{nm}^{Dr},\ f_{nm}^{C\theta}$ and $f_{nm}^{D\theta}$}\label{appb}

The expressions for the functions $f_{nm}^{Cr},\ f_{nm}^{Dr},\ f_{nm}^{C\theta}$ and $f_{nm}^{D\theta}$ appearing in Eqs.~(\ref{fincancel1}-\ref{dV21}) are:
\begin{align}\label{f1f2f3f4}
  f_{nm}^{Cr}(\widetilde{d}) &= \frac{(2 n+1)}{2}  \int_{0}^{\pi}\left[g_m^1 Y_m\left(\psi\right) +g_m^2 Y_{m+1}\left(\psi\right)\right] Y_n(\theta_1)d\theta_1 , \\
  f_{nm}^{Dr}(\widetilde{d}) &= \frac{(2 n+1)}{2}  \int_{0}^{\pi}\left[g_m^3 Y_m\left(\psi\right) +g_m^4 Y_{m+1}\left(\psi\right)\right] Y_n(\theta_1)d\theta_1 , \\
  f_{nm}^{C\theta}(\widetilde{d}) &= \frac{(2 n+1)}{2n(n+1)} \sqrt{\frac{2n+1}{4\pi}}  \int_{0}^{\pi} \left[g_m^5 Y_m \left(\psi\right) +  g_m^6 Y_{m+1}\left(\psi\right)\right] [\cos(\theta_1) P_n(\theta_1) - P_{n+1}(\theta_1)]d\theta_1 , \\
  f_{nm}^{D\theta}(\widetilde{d}) &= \frac{(2 n+1)}{2n(n+1)} \sqrt{\frac{2n+1}{4\pi}} \int_{0}^{\pi} \left[ g_m^7 Y_m \left(\psi\right) +  g_m^8 Y_{m+1}\left(\psi\right)\right] [\cos(\theta_1) P_n(\theta_1) - P_{n+1}(\theta_1)]d\theta_1 .
\end{align}
We used the identity \cite{YPn}
\begin{equation}\label{Yder}
\frac{dY_n(\theta)}{d\theta}=\sqrt{\frac{2n+1}{4\pi}}(n+1)\left[\frac{P_{n+1}(\cos \theta )}{\sin \theta }-\cot \theta  P_n(\cos \theta )\right]
\end{equation}
to rewrite the derivatives $\frac{dY_n(\theta)}{d\theta}$ in terms of $\theta$, and for the sake of brevity we defined:
\begin{align}
    g_m^1 &= \left\{\widetilde{d}^2 \left[m^2-m (3-4 \nu )-4 (1-\nu )\right]- 2 \widetilde{d} \left(m^2-2+2 \nu \right) \cos (\theta_1 )+  m^2 + m (3-4 \nu ) \right\}  \sin (\theta_1 ) / {\zeta^{{m+1}}} ,
\end{align}
\begin{align}
    g_m^2 &=-\left[\widetilde{d} (m+1)(m-4+4 \nu )\right] \sin (\theta_1 ) /{\zeta^{m}} ,
\end{align}
\begin{align}
    g_m^3 &=-\left( m+1\right) \sin (\theta_1 ) / {\zeta^{m+1}} ,
\end{align}
\begin{align}
    g_m^4 &=\widetilde{d} (m+1) \sin (\theta_1) / {\zeta^{m+2}} ,
\end{align}
\begin{align}
    g_m^5 &= \frac{-\left(\widetilde{d}^2+1\right) (m+1) (m-4+4 \nu ) \cos (\theta_1 )+ \widetilde{d} \left[ m (m-6+8 \nu )-6 (1-\nu )+ \left(m^2-2+2 \nu \right) \cos (2 \theta_1 )\right]}{\sin(\theta_1) \zeta^{m+2}} ,
\end{align}
\begin{align}
    g_m^6&=-(m+1) (m-4+4 \nu ) / [\sin(\theta_1){\zeta^{m}}] ,
\end{align}
\begin{align}
    g_m^7&=(m+1) \cot (\theta_1) / {\zeta^{m+1}} ,
\end{align}
\begin{align}
    g_m^8&=-[\widetilde{d} \cos (\theta_1 )-1] (m+1) / [\sin (\theta_1){\zeta^{m+2}}] ,
\end{align}
where
\begin{align}\label{psizeta}
  \psi &\equiv \left[ \widetilde{d}-\cos (\theta_1) \right]/{\zeta} , \\
  \zeta &\equiv \sqrt{d^2-2d\cos(\theta_1)+1}.
\end{align}

\section{The stress tensor}\label{appc}

Here we develop the expressions for the stress tensor in the cases discussed in the paper. The applied displacements are symmetric about an axis passing through the centers of the two force dipoles. The expressions are taken from~\cite{01} for the case of the displacement field given by Eqs.~(\ref{uri}-\ref{uti}), excluding the first term in Eq.~(\ref{uri}) which corresponds to volume change. The stress tensor $\underline{\underline{\tau}}$ may be written in the following form in this case:
\begin{equation}\label{taudef}
  \underline{\underline{\tau}}=\sum_{n=0}^{\infty} \left(
                                                     \begin{array}{ccc}
                                                       \tau_{rr}^{(n)} & \tau_{r \theta}^{(n)} & \tau_{r \varphi}^{(n)} \\
                                                       \tau_{\theta r}^{(n)} & \tau_{\theta \theta}^{(n)} & \tau_{\theta \varphi}^{(n)} \\
                                                       \tau_{\varphi r}^{(n)} & \tau_{\varphi \theta}^{(n)} & \tau_{\varphi \varphi}^{(n)} \\
                                                     \end{array}
                                                   \right)
\end{equation}
The stress tensor is symmetric $\widetilde{\tau}_{ij}=\widetilde{\tau}_{ji}$ and thus only six components are to be evaluated. We define the dimensionless stress tensor $\underline{\underline{\tau}}=\frac{\tau}{G} \frac{R_0}{u_0}$, the elements of which are given by:
\begin{align}
  \widetilde{\tau}_{RR}^{(n)} = &2  \left[ - \frac{C_n}{\widetilde{r_i}^{n+1}} n (n^2+3n-2 \nu) + \frac{D_n}{\widetilde{r_i}^{n+3}}(n+1)(n+2) \right] Y_n \left(cos{\theta_i} \right)\ \ \  \label{sig1} , \\
  \widetilde{\tau}_{R \theta}^{(n)} = &2 \left[ \frac{C_n}{\widetilde{r_i}^{n+1}} (n^2-2+2 \nu) - \frac{D_n}{\widetilde{r_i}^{n+3}}(n+2) \right] \frac{dY_n\left(cos{\theta_i} \right)}{d \theta_i} \label{sig2} , \\
  \widetilde{\tau}_{\theta \theta}^{(n)} = &2 \left\{ \left[\frac{C_n}{\widetilde{r_i}^{n+1}} n (n^2-2n-1+2 \nu) - \frac{D_n}{\widetilde{r_i}^{n+3}}(n+1)^2 \right] Y_n(cos{\theta_i})
    -  \left[ \frac{C_n}{\widetilde{r_i}^{n+1}} (-n+4-4 \nu) + \frac{D_n}{\widetilde{r_i}^{n+3}} \right] \frac{dY_n(cos{\theta_i})}{d \theta_i} ctg{\theta} \right\} , \label{sig3} \\
  \widetilde{\tau}_{\varphi \varphi}^{(n)} =  &2 \left\{ \left[ \frac{C_n}{\widetilde{r_i}^{n+1}} n (n+3-4n \nu-2 \nu) -  \frac{D_n}{\widetilde{r_i}^{n+3}}(n+1) \right] Y_n(cos{\theta_i})
    + \left[ \frac{C_n}{\widetilde{r_i}^{n+1}} (-n+4-4 \nu) + \frac{D_n}{\widetilde{r_i}^{n+3}} \right] \frac{dY_n(cos{\theta_i})}{d \theta_i} ctg{\theta} \right\} , \label{sig4} \\
  \widetilde{\tau}_{R \varphi}^{(n)} = &\widetilde{\tau}_{\theta \varphi}^{(n)}=0 . \label{sig5}
\end{align}

\section{Derivation of Eq.~(\ref{Eint})}\label{appe}

Here we evaluate the integral~(\ref{Eint}) for the interaction energy of two ``live'' spherical force dipoles. Without loss of generality, we take the initial isotropic displacement in the positive direction:
\begin{align}\label{ufs}
&2 \cdot \frac{1}{2} \int_S \overrightarrow{u} \cdot \overrightarrow{F} ds =
\left. 2 \frac{1}{2} \int_0^{\pi} \left[ (u_r F_r)+(u_{\theta} F{_{\theta}}) \right] 2\pi R_0^2 \sin{\theta} d\theta \right|_{\widetilde{r}=1} \nonumber\\
&=-\left. 2\pi u_0^2 G R_0 \int_{-1}^{1} \left\{ \left(  - \frac{1}{\widetilde{r}^2} +\widetilde{u}_{r11}+\widetilde{u}_{r21}\right) \left[-\left(\frac{4}{\widetilde{r}^{3}}+ \underline{\underline{\tau}}_{11}\right) \right] +\left(\widetilde{u}_{\theta11}+\widetilde{u}_{\theta21}\right) \left( - \underline{\underline{\tau}}_{22}\right) \right\}  d cos {\theta} \right|_{\widetilde{r}=1}
\end{align}
Here $\widetilde{u}_{r11},\ \widetilde{u}_{r21},\ \widetilde{u}_{\theta11},\ \widetilde{u}_{\theta21}$ and $\underline{\underline{\tau}}$ are defined below [see Eqs.~(\ref{surfr1}-\ref{surft2}) and (\ref{taudef}-\ref{sig5})]. The terms $-\frac{1}{\widetilde{r}^2}$ and $-\frac{4}{\widetilde{r}^2}$ are the isotropic self-displacement created by each spherical force dipole and the corresponding force applied on it in response by the environment. Noting the fact that all the terms in~(\ref{ufs}) are given in terms of Legendre polynomials $P_n(cos \theta)$ and their derivatives $\frac{dP_n(cos \theta)}{d \theta}$ and using their orthogonality, after integration and cancellation of the appropriate terms we get:
\begin{align}\label{ufs2}
&16\pi u_0^2 G R_0 \left( 1 + D_0 - \left\{ - D_0 + \sum_{m=0}^{\infty} \left[ f_{0m}^{Cr} C_m + f_{0m}^{Dr} (D_m - \sqrt{4\pi}\delta_{m,0}) \right] \right. \right\}  \nonumber \\
&- \sum_{n=1}^{\infty} \left\{n (n+3-4 \nu) {C_n} - (n+1) {D_n} + \sum_{m=0}^{\infty} \left[ f_{nm}^{Cr} C_m + f_{nm}^{Dr} (D_m - \sqrt{4\pi}\delta_{m,0}) \right] \right\} \nonumber\\
&\cdot \left[ n (n^2+3n-2 \nu) {C_n} - (n+1)(n+2) {D_n} \right] \cdot \frac{2}{4(2n+1)} \nonumber \\
&- \sum_{n=1}^{\infty} \left\{ (-n+4-4 \nu) {C_n}  + {D_n}  + \sum_{m=0}^{\infty} \left[ f_{nm}^{C\theta} C_m + f_{nm}^{D\theta} (D_m - \sqrt{4\pi}\delta_{m,0}) \right] \right\} \nonumber \\
&\cdot 2 \left[- (n^2-2+2 \nu) {C_n} + (n+2){D_n} \right] \frac{n (n+1)}{2n+1} \Bigg)
\end{align}

In the FP cases the coefficients $C_n$, $D_n$ for any $n$ were either evaluated using the appropriate cancellation condition, Eqs. (\ref{surf1}-\ref{surf2}) or assumed to be zero. In both FP cases the last two terms in Eq. (\ref{ufs}) vanish and the result becomes:
\begin{align}\label{ufsres}
  2 \cdot \frac{1}{2} \int_S \overrightarrow{u} &\cdot \overrightarrow{F} ds = 16\pi u_0^2 G R_0 \left( 1 + D_0 - \left\{ - D_0 + \sum_{m=0}^{\infty} \left[ f_{0m}^{Cr} C_m + f_{0m}^{Dr} (D_m - \sqrt{4\pi}\delta_{m,0}) \right] \right\} \right)
\end{align}

In the VP cases on the other hand we do not calculate the coefficients $C_1$ and $D_1$ using the cancellation conditions, but using the conditions (\ref{c1d11}-\ref{c1d12}) instead. Thus in VP cases neither the displacements, nor the stress vanish for $n=1$. Using (\ref{c1d11}-\ref{c1d12}) we get in these cases:
\begin{align}\label{ufsres2}
  2 \cdot \frac{1}{2} \int_S \overrightarrow{u} \cdot \overrightarrow{F} ds &= 16\pi u_0^2 G R_0 \left( 1 + D_0 - \left\{ - D_0 + \sum_{m=0}^{\infty} \left[ f_{0m}^{Cr} C_m + f_{0m}^{Dr} (D_m - \sqrt{4\pi}\delta_{m,0}) \right] \right\} \right. \nonumber \\
  & {} - \frac{1}{6}\left[(4-2\nu)C_1-6 D_1\right]\sum_{m=0}^{\infty} \left( \frac{f_{1m}^{C\theta}-f_{1m}^{Cr}}{2} C_m + \frac{f_{1m}^{D\theta} - f_{1m}^{Dr}}{2} D_m \right) \nonumber \\
  & \left. {} - \frac{4}{3}\left[(1-2\nu)C_1+3 D_1\right]\sum_{m=0}^{\infty} \left( \frac{f_{1m}^{Cr}-f_{1m}^{C\theta}}{2} C_m + \frac{f_{1m}^{Dr}-f_{1m}^{D\theta}}{2} D_m \right) \right)
\end{align}

The last two terms in Eq.~(\ref{ufsres2}) are proportional to higher powers of $C_n$ and $D_n$. It may be seen from Table~\ref{tabcoef} that the coefficients $C_n$ and $D_n$ are inversely proportional to $\widetilde{d}$, and thus at large distances the last two terms in Eq.~(\ref{ufsres2}) may be neglected and Eq.~(\ref{ufsres2}) reduces to Eq.~(\ref{ufsres}).

\section{$D_0$, $C_1$, $D_1$, $C_2$, $D_2$ and $\Delta E$ for arbitrary distance}\label{appf}

In Table~\ref{tabcoef2} we present the expressions for the coefficients $D_0$, $C_1$, $D_1$, $C_2$, $D_2$ and for the dimensionless interaction energy $\Delta \widetilde{E}$ without the large-distance assumption that appears in Table~\ref{tabcoef}. 

\begin{table}[h]
\begin{center}
\begin{tabular}{| c | c | c | c | c | c | c | c |}
    \hline
      & Fixed Size & Fixed Size & Variable Size & Variable Size \\
      & Fixed Position & Variable Position & Fixed Position & Variable Position \\
      & ($FSFP$) & ($FSVP$) & ($VSFP$) & ($VSVP$) \\  \hline
      $n_{max}$ & 1 & 2 & 1 & 2 \\ \hline
    $D_0$ & $\frac{5 (1-2 \nu) \widetilde{d}^2}{\Lambda_1}$ & $\frac{\lambda_1}{\Lambda_2}$ & 0 & 0 \\[1.3ex] \hline
    $C_1$ & $-\frac{15 \widetilde{d}^4}{2 \Lambda_1}$ & $\frac{315 (2-3\nu) \widetilde{d}^6}{\Lambda_2}$ & $-\frac{15 \widetilde{d}^4}{2 \Lambda_3}$ & $\frac{63 (2-3 \nu ) \widetilde{d}^3}{\Lambda_4} $\\[1.3ex] \hline
    $D_1$ & $-\frac{ \widetilde{d} [5 \widetilde{d}^3 -(8 -12\nu)]}{2 \Lambda_1}$ & $\frac{315 (2 - 3 \nu) (7 - 8 \nu) \widetilde{d}^6}{\Lambda_2}$ & $-\frac{ \widetilde{d} [5 \widetilde{d}^3 -(8 -12\nu)]}{2 \Lambda_3}$ & $\frac{63 (2 - 3 \nu) (7 - 8 \nu) \widetilde{d}^3}{\Lambda_4}$ \\ [1.3ex]\hline
    $C_2$ & 0 & $-\frac{35 \left[5 (5-6 \nu ) \widetilde{d}^3+(2-3 \nu)\right]\widetilde{d}^7}{\Lambda_2}$ & 0 & $-\frac{35 (5-6 \nu ) \widetilde{d}^7}{\Lambda_4}$ \\[1.3ex] \hline
    $D_2$ & 0 & $-\frac{\lambda_2}{\Lambda_2}$ & 0 & $-\frac{6 \left(7 \widetilde{d}^5-28+40 \nu \right) \widetilde{d}^2}{\Lambda_4}$ \\[1.3ex] \hline
    $\Delta \widetilde{E}$ & $\frac{10 (1-2 \nu) \widetilde{d}^2}{\Lambda_1}$ & $\frac{\lambda_3}{\Lambda_2^2}$ & $-\frac{10 (1-2 \nu)
\widetilde{d}^2}{\Lambda_3}$ & $-\frac{\lambda_4}{\Lambda_4}$ \\ [1.3ex] \hline
$\Delta \widetilde{E}_{\infty}$ & $\frac{2 (1-2 \nu)
}{(5-6\nu)}\cdot \frac{1}{\widetilde{d}^{4}}$ & $\frac{10 (1-2 \nu)}{(4-5\nu)} \cdot \frac{1}{\widetilde{d}^{6}}$ & $-\frac{2 (1-2 \nu)
}{(5-6\nu)} \cdot \frac{1}{\widetilde{d}^{4}}$ & $-\frac{10 (1-2 \nu) }{(4-5\nu)}\cdot \frac{1}{\widetilde{d}^{6}}$ \\ \hline
\end{tabular}
\caption{Coefficients $C_n$ and $D_n$ of the first terms of the multipole expansion. $n_{max}$ denotes the highest-order term taken into account, $\Delta \widetilde{E} \equiv \frac{E_{int}}{E_0}$ is the resultant dimensionless interaction energy and $\Delta \widetilde{E}_{\infty}$ is its asymptotic long-distance behavior given also in Table~\ref{tabcoef}. $\widetilde{d}=\frac{d}{R_0}$ is the dimensionless distance between the force dipoles.} \label{tabcoef2}
\end{center}
\end{table}

We use the following expressions:
\begin{align}
  \Lambda_1 &= 5  (5-6 \nu) \widetilde{d}^6 - 30  (1 - \nu) \widetilde{d}^5 + 10 \widetilde{d}^3 - (5 - 10 \nu) \widetilde{d}^2 - (8   - 12 \nu) \label{L1} ,
\end{align}
\begin{align}
  \Lambda_2 &=140 \left(20-49 \nu + 30 \nu ^2\right)\widetilde{d}^{13}- 28 \left(242-403 \nu + 135 \nu ^2 \right) \widetilde{d}^{10} +
            2520 (5-6 \nu) \widetilde{d}^8 -35\left(86-283 \nu + 216 \nu ^2 \right) \widetilde{d}^7 \nonumber \\
            &{}-1260 \left(10-27 \nu + 18 \nu ^2 \right)\widetilde{d}^5+70 \left(2-7 \nu +6 \nu ^2 \right) \widetilde{d}^4 -720 \left(35-92 \nu  +60 \nu ^2\right) \widetilde{d}^3 -36 \left(14-41 \nu +30 \nu ^2\right) \label{L2} ,
\end{align}
\begin{align}
  \Lambda_3 &=5 (5-6 \nu) \widetilde{d}^6  - 30 (1 - \nu) \widetilde{d}^5 + 10 \widetilde{d}^3  - (8   - 12 \nu) \label{L3} ,
\end{align}
\begin{align}
  \Lambda_4 &=4 (5-6 \nu ) \left[7 (4-5 \nu ) \widetilde{d}^{10}-35 (2-\nu) \widetilde{d}^7 + 126 \widetilde{d}^5-36 (7-10 \nu)\right] \label{L4} ,
\end{align}
\begin{align}
  \lambda_1 &=70 (1-2 \nu ) \left[10 (5-6 \nu ) \widetilde{d}^3-(2-3\nu)\right] \widetilde{d}^4 ,
\end{align}
\begin{align}
  \lambda_2 &=\left[35 \widetilde{d}^8 (5-6 \nu ) + 7 \widetilde{d}^5 (2-3 \nu )- 20 \widetilde{d}^3 \left(35-92 \nu +60 \nu^2\right)- \left(14-41 \nu +30 \nu ^2\right) \right] 6 \widetilde{d}^2,
\end{align}
\begin{align}
  \lambda_3 &=70\widetilde{d}^4 \left[ 2800 (5-6 \nu )^2 \left(10\nu ^2-13 \nu +4\right) \widetilde{d}^{16}+ 420 \left(330 \nu ^4+1411 \nu ^3-3961 \nu^2+2898 \nu -640\right) \widetilde{d}^{13} \right. \nonumber\\
            &{}- 50400 (5-6 \nu )^2 (2 \nu -1) \widetilde{d}^{11}-14\left(40230 \nu ^4-81594 \nu ^3+50371 \nu ^2-6691 \nu -1586\right) \widetilde{d}^{10} \nonumber\\
            &{}-630\left(7074 \nu ^4-17937 \nu ^3+15963 \nu ^2-5636 \nu +580\right) \widetilde{d}^8-105\left(3396 \nu ^4-11879 \nu ^3+15041 \nu ^2-8232 \nu +1652\right) \widetilde{d}^7 \nonumber\\
            &{}-14400(5-6 \nu )^2 \left(20 \nu ^2-24 \nu +7\right) \widetilde{d}^6-504 (2-3 \nu )^2 \left(375\nu ^2-643 \nu +275\right) \widetilde{d}^5 - 140 \left(6 \nu ^2-7 \nu +2\right)^2 \widetilde{d}^4 \nonumber\\
            &{}-360\left(14940 \nu ^4-46008 \nu ^3+52895 \nu ^2 \left.  -26907 \nu +5110\right) \widetilde{d}^3-18 (2-3 \nu )^2 \left(790 \nu ^2-1263 \nu +497\right) \right] ,
\end{align}
\begin{align}
  \lambda_4 &=28 \widetilde{d} \left[ 1400 (5-6 \nu )^2 \left(4-13 \nu +10 \nu ^2\right)\widetilde{d}^{13}
            -70 \left(8000-39830 \nu +71471 \nu ^2-54759 \nu ^3+15030\nu ^4\right) \widetilde{d}^{10}  \right. \nonumber \\
            &{} + 25200 (5-6\nu )^2 (1-2 \nu ) \widetilde{d}^8 - 525 \left(250-1465 \nu +3163 \nu^2-2988 \nu ^3+1044 \nu^4\right) \widetilde{d}^7 \nonumber \\ &{} - 945 \left(500-2180\nu +3431 \nu ^2-2277 \nu ^3+522 \nu ^4\right) \widetilde{d}^5 + 1890 (2-3 \nu)^2 \left(25-54 \nu +29 \nu^2\right) \widetilde{d}^4 \nonumber \\
            &{} - 7200 (5-6 \nu )^2\left(7-24 \nu +20 \nu ^2\right)\widetilde{d}^3 - 567 (2-3 \nu )^2\left(300-673 \nu +377 \nu^2\right) \widetilde{d}^2 \nonumber \\
            &{} + \left. 540\left(1750-9255 \nu +18281 \nu^2-15984 \nu ^3+5220 \nu^4\right)  \vphantom{\widetilde{d}} \right] .
\end{align}

\bibliography{two_spheres_njp}

\end{document}